\documentclass[11pt, a4paper]{article}
\usepackage{moriond,epsfig,amsmath,wrapft}
\begin{document}
\vspace*{4cm}
\title{Global analysis for determining fragmentation functions \\
        and their uncertainties in light hadrons}
\author{ M. Hirai (1), S. Kumano (2,3), T.-H. Nagai (3), K. Sudoh (2)}
\address{(1) Department of Physics,
             Tokyo Institute of Technology \\
             Ookayama, Meguro-ku, Tokyo, 152-8550, Japan \\
         (2) Institute of Particle and Nuclear Studies, 
             High Energy Accelerator Research Organization (KEK) \\
             1-1, Ooho, Tsukuba, Ibaraki, 305-0801, Japan \\
         (3) Department of Particle and Nuclear Studies,
             Graduate University for Advanced Studies \\
             1-1, Ooho, Tsukuba, Ibaraki, 305-0801, Japan}     
\maketitle
\abstracts{
Fragmentation functions are determined for the pion, kaon, and proton
by analyzing charged-hadron production data in electron-positron
annihilation. It is important that uncertainties of
the determined fragmentation functions are estimated in this analysis.
Analysis results indicate that gluon and light-quark functions have 
large uncertainties especially at small $Q^2$. We find that 
next-to-leading-order (NLO) uncertainties are significantly reduced
in comparison with leading-order (LO) ones in the pion and kaon.
The fragmentation functions are very different in various analysis groups.
However, all the recent functions are roughly within the estimated
uncertainties, which indicates that they are consistent with each other.
We provide a code for calculating the fragmentation functions and their
uncertainties at a given kinematical point of $z$ and $Q^2$ by a user.}
\noindent
{\small¥{\it Keywords}: Fragmentation function, Quark, Gluon, QCD,
                        Electron-positron annihilation} 

\section{Introduction}\label{intro}

A fragmentation function describes a hadronization process from a parent
quark, antiquark, or gluon to a hadron. Hadron-production processes
are often used for investigating important physics such as
the origin of nucleon spin and properties of quark-hadron matters.
Fragmentation functions are needed for describing such processes,
so that precise functions should be obtained
for discussing any physics outcome. Nevertheless,
it is known that there are large differences in the parametrized
fragmentation functions, for example, between the ones by
Kniehl, Kramer, and P\"otter (KKP) \cite{kkp} and Kretzer \cite{kretzer}. 
Recently updated functions by Albino, Kniehl, and Kramer (AKK) \cite{akk}
are also much different from these functions. This fact suggests that
the fragmentation functions are not determined accurately;
therefore, it is important to show reliable regions in discussing
any hadron-production data.

Such error analyses have been investigated recently in the studies of
unpolarized parton distribution functions (PDFs), polarized and
nuclear PDFs \cite{pdf-error}. It is straightforward to apply
the technique for the fragmentation functions.
We determine the fragmentation functions and their uncertainties by
analyzing the data for charged-hadron production in electron-positron
annihilation, $e^+ + e^- \rightarrow h+X$. The analyses are done
in leading order (LO) and next-to-leading order (NLO) of the
running coupling constant $\alpha_s$. Because accurate SLD data in 2004 
are included in our analysis, whereas they are not used in KKP, AKK,
and Kretzer's analyses, we expect to have improvements.
Therefore, important points of our analysis are \cite{hkns07}
\begin{itemize}
\vspace{-0.15cm}
\item improvement due to addition of accurate SLD data,
\vspace{-0.15cm}
\item roles of NLO terms on the determination, namely on the uncertainties,
\vspace{-0.15cm}
\item comparison with other analysis results by considering
      the uncertainties.
\end{itemize}
Our analysis method is explained in section 2,  
results are explained in section 3, and they are summarized in section 4.

\section{Analysis method}

The fragmentation function is defined by the ratio of hadron-production
cross section to the total hadronic cross section:
\begin{equation}  
F^h(z,Q^2) = \frac{1}{\sigma_{tot}} 
\frac{d\sigma (e^+e^- \rightarrow hX)}{dz} ,
\end{equation}
where $Q^2$ is given by the center-of-mass energy squared ($Q^2=s$),
and $z$ is defined by the ratio
$z = E_h/(\sqrt{s}/2) = 2E_h/\sqrt{Q^2}$ with the hadron energy $E_h$.
Since the fragmentation occurs from primary quarks, antiquarks,
and gluons, the fragmentation function is expressed by the their sum:
\begin{equation}  
F^h(z,Q^2) = \sum_i \int^{1}_{z} \frac{dy}{y}
C_i(y,\alpha_s)  D_i^h (z/y,Q^2) .
\end{equation}
Here, $C_i(z,\alpha_s)$ is a coefficient function which is calculated
in perturbative QCD, and $D_i^h(z,Q^2)$ is the fragmentation function
of the hadron $h$ from a parton $i$. The function $D_i^h(z,Q^2)$ is
associated with a non-perturbative aspect, and it cannot be 
theoretically calculated in a reliably way. It is the purpose of
this work to obtain the optimum fragmentation functions for the pion,
kaon, and proton by analyzing the experimental data for
$e^+ +e^- \rightarrow h+X$.

In order to determine the functions from the data, we express them
in terms of parameters at a fixed scale $Q_0^2$ (=1 GeV$^2$):
\begin{equation}
D_i^h(z,Q_0^2) = N_i^h z^{\alpha_i^h} (1-z)^{\beta_i^h} ,
\end{equation}
where $N_i^h$, $\alpha_i^h$, and $\beta_i^h$ are the parameters
to be determined by a $\chi^2$ analysis of the data.
Because there is a sum rule due to the energy conservation:
$\sum_h M_i^h = \sum_h \int_0^1 dz \, z \, D_i^h (z,Q^2) = 1$,
it is more convenient to choose the parameter $M_i^h$ instead
of $N_i^h$. They are related by 
$N_i^h = M_i^h /B(\alpha_i^h+2, \beta_i^h+1)$, where 
$B(\alpha_i^h+2, \beta_i^h+1)$ is the beta function.
In general, a common function is assumed for favored functions
and different ones are used for disfavored functions.
The favored indicates a fragmentation from a quark or antiquark
which exists in the hadron as a constituent in a simple quark model.
The disfavored means a fragmentation from other quark or
antiquark. The details of the formalism are explained in Ref. 5.
The optimum parameters are determined by minimizing the total
$\chi^2$ given by
$\chi^2 = \sum_j (F_{j}^{data}-F_{j}^{theo})^2 /(\sigma_j^{data})^2$,
where $F_{j}^{data}$ and $F_{j}^{theo}$ are experimental and
theoretical fragmentation functions, respectively, and
$\sigma_j^{data}$ is an experimental error.
Uncertainties of the determined fragmentation functions are estimated
by the Hessian method:
\begin{equation}
[\delta D_i^h (z)]^2=\Delta \chi^2 \sum_{j,k}
\left( \frac{\partial D_i^h (z,\xi)}{\partial \xi_j}  \right)_{\hat\xi}
H_{jk}^{-1}
\left( \frac{\partial D_i^h (z,\xi)}{\partial \xi_k}  \right)_{\hat\xi}
\, ,
\end{equation}
where $H_{jk}$ is the Hessian matrix, $\xi_j$ is a parameter,
$\hat\xi$ indicates the optimum parameter set, and the $\Delta \chi^2$
value is chosen so that the error becomes the one-$\sigma$ range in
the multiparameter space. The detailed explanations for
the uncertainties are found in Refs. 4 and 5.

\section{Results}

\vspace{0.0cm}
\begin{wrapfigure}{r}{0.35\textwidth}
   \vspace{-0.3cm}
   \begin{center}
       \epsfig{file=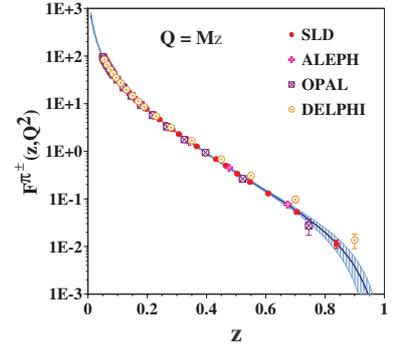,width=5.0cm} \\
       \vspace{-0.3cm}
\caption{NLO results are compared with charged-pion data.}
\label{fig:ff-data-pi}
   \end{center}
\end{wrapfigure}
\vspace{0.0cm}

We explain analysis results. First, determined fragmentation functions
are compared with charged-pion production data in 
Fig. \ref{fig:ff-data-pi}  \cite{hkns07}.
The curve indicates theoretical NLO results which are calculated by
using determined parameters in the $\chi^2$ analysis, and the uncertainties
are shown by the shaded band. The comparison suggests that the fit
is successful in reproducing the data in four orders of magnitude. 

\begin{figure}[h!]
\parbox[t]{0.50\textwidth}{
   \begin{center}
       \vspace{-0.1cm}
       \epsfig{file=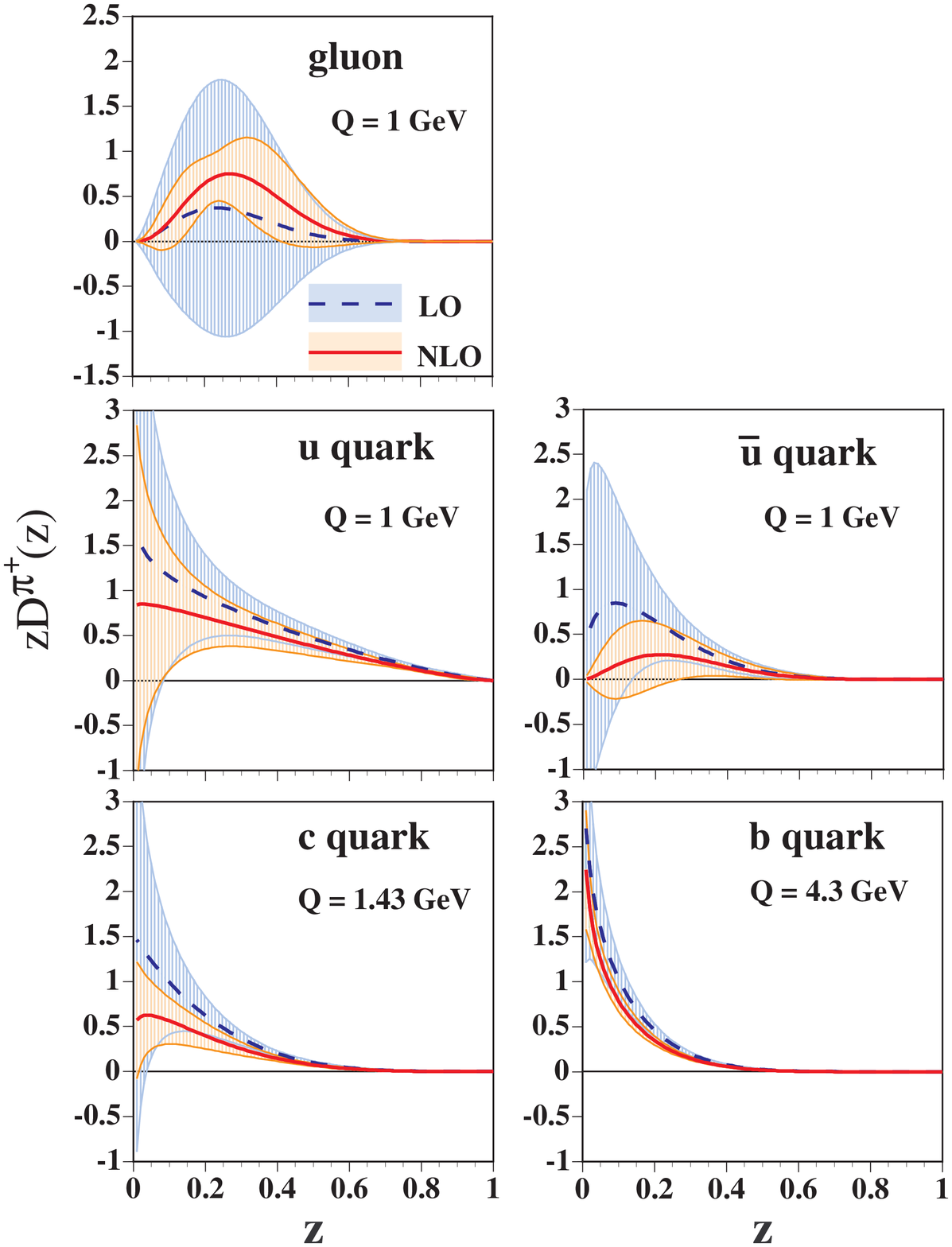,width=5.5cm} \\
   \end{center}
}
\parbox[t]{0.50\textwidth}{
   \begin{center}
       \vspace{-0.1cm}
       \epsfig{file=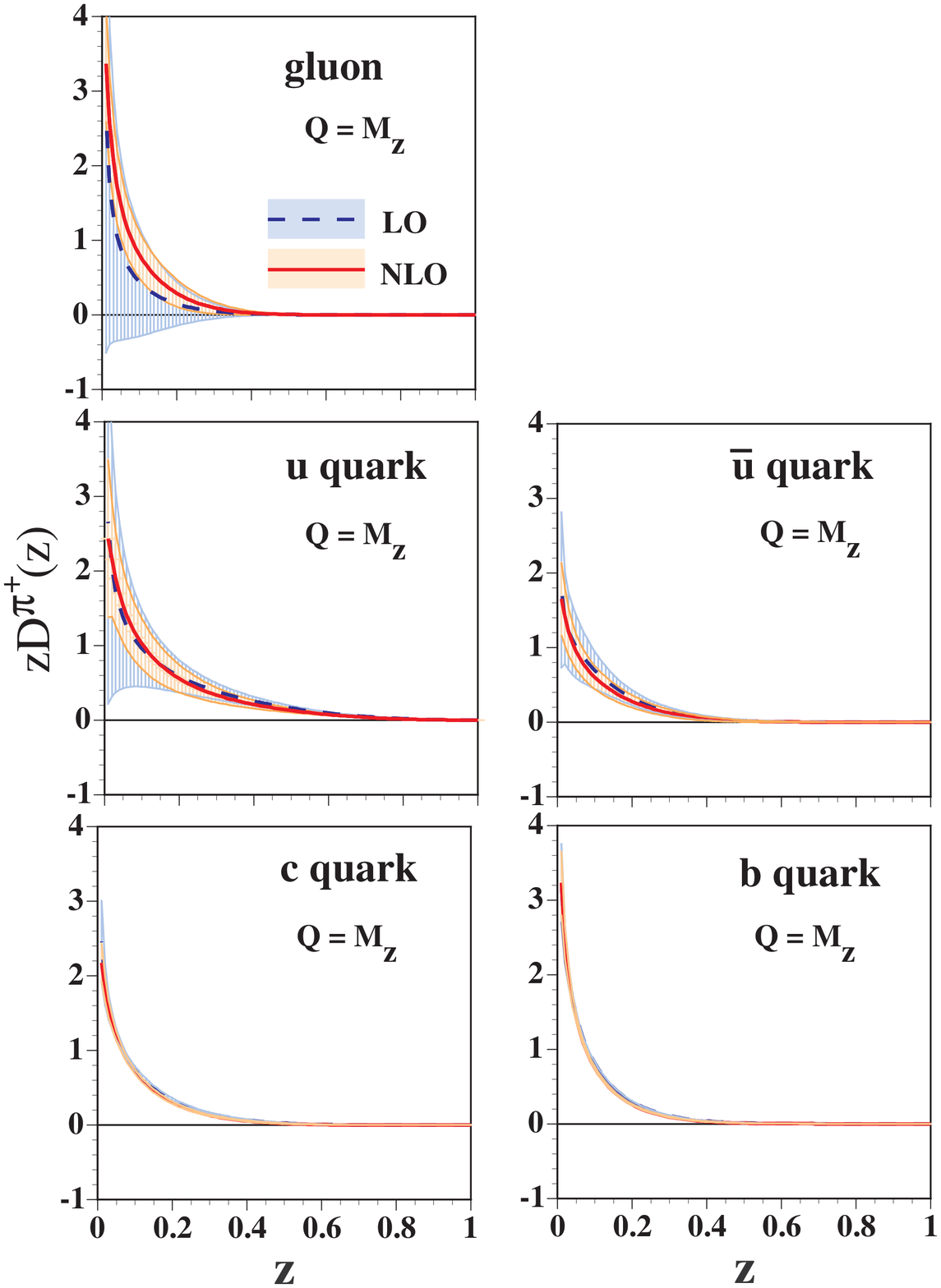,width=5.5cm} \\
   \end{center}
}
\vspace{-0.3cm}
\caption{Determined fragmentation functions for $\pi^+$, $K^+$,
         and proton at $Q^2=1$ GeV$^2$, $m_c^2$, $m_b^2$, and $M_Z^2$.
         LO and NLO functions are shown with their uncertainties.}
\label{fig:ff-pion}
\end{figure}

Determined functions are shown at the initial scales ($Q^2$=1 GeV$^2$,
$m_c^2$, and $m_b^2$) and also at an evolved scale $Q^2=M_Z^2$ in Fig.
\ref{fig:ff-pion} \cite{hkns07}.
The LO and NLO functions and their uncertainties
are shown. We notice that the uncertainties are generally large
at small $Q^2$, especially in the LO. 
The gluon and light-quark functions have especially large
uncertainties. However, it is interesting to note that the situation
is much improved in the NLO because the uncertainties become
significantly smaller. The uncertainty bands are smaller
at large $Q^2$ $(=M_Z^2)$. Since the fragmentation functions are used
at small $Q^2$ ($\sim 1$ GeV$^2$), for example, in HERMES, RHIC-Spin,
and RHIC heavy-ion experiments, one should be careful about
the reliability of employed functions in one's analysis.

\noindent

\begin{figure}[t!]
\parbox[t]{0.33\textwidth}{
   \begin{center}
       \epsfig{file=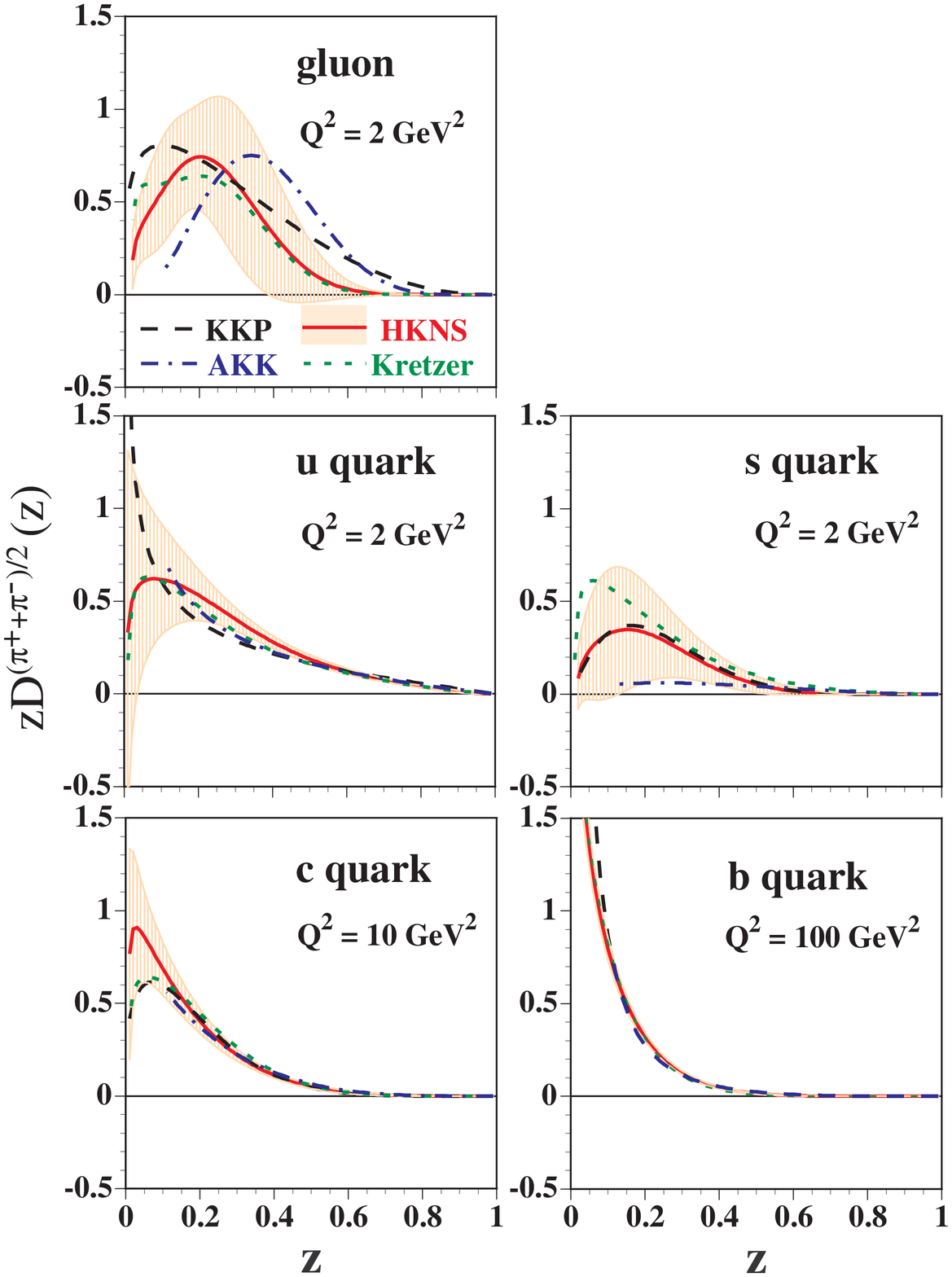,width=5.0cm} \\
   \end{center}
}\hfill
\parbox[t]{0.33\textwidth}{
   \begin{center}
       \vspace{-0.1cm}
       \epsfig{file=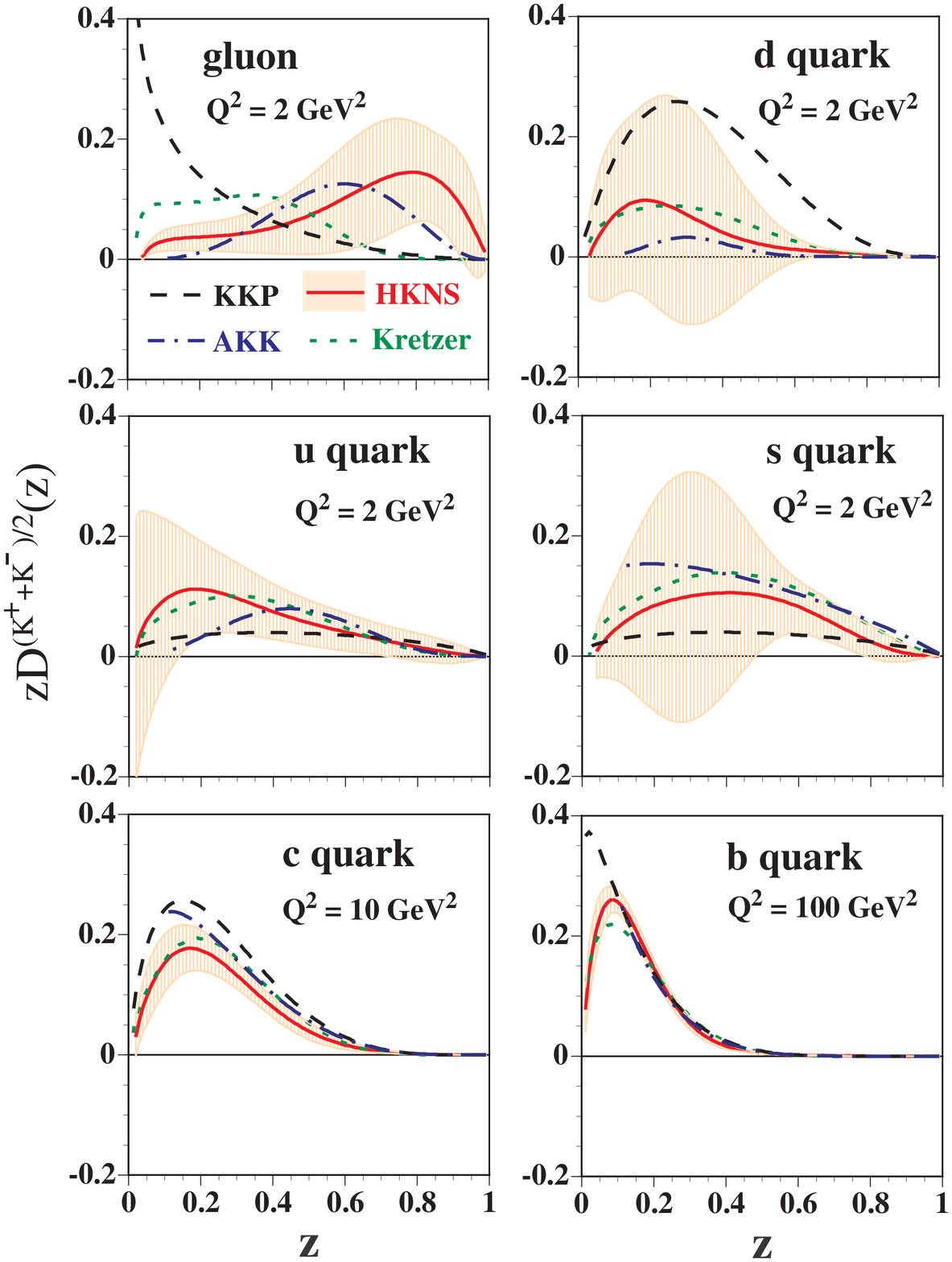,width=5.0cm} \\
   \end{center}
}
\parbox[t]{0.33\textwidth}{
   \begin{center}
       \vspace{-0.1cm}
       \epsfig{file=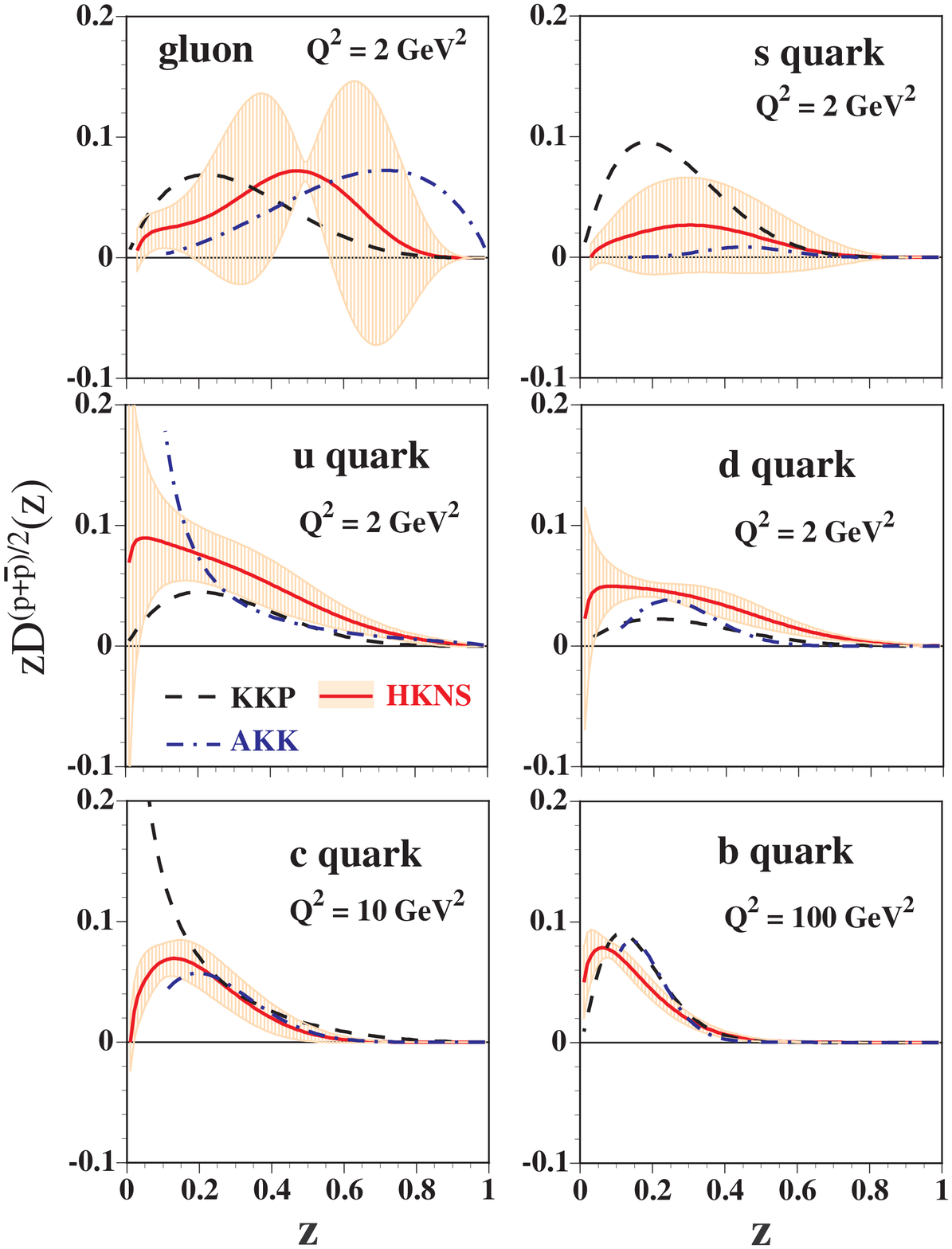,width=5.0cm} \\
   \end{center}
}
\vspace{-0.3cm}
\caption{Fragmentation functions for 
         $(\pi^+ +\pi^-)/2$, $(K^+ +K^-)/2$, and $(p+\bar p)/2$
         are compared with other analysis results 
         at $Q^2$=2, 10, and 100 GeV$^2$.}
\label{fig:ffs-others}
\end{figure}

Next, the determined functions are compared with other analysis results
for $(\pi^+ +\pi^-)/2$, $(K^+ +K^-)/2$, and $(p+\bar p)/2$
in Fig. \ref{fig:ffs-others} \cite{hkns07}.
Our parametrization is denoted HKNS
(Hirai, Kumano, Nagai, Sudoh). The determined functions in NLO and
their uncertainties are shown by the solid curves and shaded bands.
They are compared with other functions by KKP, AKK, and Kretzer
at $Q^2$=2, 10, and 100 GeV$^2$. As mentioned earlier, there are
much differences between the analysis groups. For example,
the gluon and $s$-quark functions have large variations in the pion.
However, almost all the curves are roughly within the estimated
uncertainty bands. It suggests that all the analyses should be
consistent with each other and that accurate functions cannot 
be determined by the current $e^+e^-$ data.
After our paper \cite{hkns07}, there appeared another analysis by
de Florian, Sassot, and Stratmann \cite{dss}. Although there are some
differences from our functions, they are also within the uncertainty
bands in Fig. \ref{fig:ffs-others}.

\noindent
\vspace{-0.50cm}

The determined fragmentation functions can be calculated by using
a code at our web site \cite{web} by supplying a kinematical
condition for $z$ and $Q^2$ and a hadron species. It is noteworthy
that the uncertainties can be also calculated by using the code.

\section{Summary}

The optimum fragmentation functions and their uncertainties have been
obtained for the pion, kaon, and proton in both LO and NLO of $\alpha_s$ 
by the $\chi^2$ analyses of charged-hadron production data in
electron-positron annihilation. It is the first analysis to show the
uncertainties in the fragmentation functions. The uncertainties 
were estimated by the Hessian method. We found large
uncertainties especially at small $Q^2$, so that they need to be taken
into account for using the functions in the small $p_T$ regions
of hadron-production measurements in lepton-proton, proton-proton,
and heavy-ion reactions. We also found that the functions are
determined more accurately in the NLO than the LO ones particularly
in the pion by considering LO and NLO uncertainties. There are large
differences between previous parametrizations of KKP, AKK, and Kretzer,
but they are consistent with each other and with our results because
they are within the uncertainty bands. 


\end{document}